\documentclass[12pt,aps,epsf]{revtex4}
\usepackage{epsf}

\usepackage{amsmath}

\begin{document}

\title{Application of Volume diffraction grating for terahertz lasing in Volume FEL (VFEL)}

\author{V.G.Baryshevsky}
\email[E-mail me at:]{bar@inp.minsk.by}
\author{K.G.Batrakov}
\email[E-mail me at:]{batrakov@inp.minsk.by}
\author{V.I.Stolyarsky}
\email [E-mail me at:] {vist@inp.minsk.by}
\affiliation{Institute
of Nuclear Problems, Belarusian State University, 11 Bobruiskaya
Str., Minsk 220050, Belarus.}

\date{\today}
\begin{abstract}
The generation of induced radiation in volume resonator formed by metal
threads is considered. It is shown that using of such volume diffraction
grating allows increasing of lasing efficiency in terahertz range. The
requirements on beam and grating parameters are obtained.
\end{abstract}
\maketitle
\section{Introduction}
\qquad  Generation of radiation in millimeter and far-infrared
range with nonrelativistic and low-relativistic electron beams
gives rise difficulties. Gyrotrons and cyclotron resonance
facilities are used as sources in millimeter and sub-millimeter
range, but for their operation magnetic field about several tens
of kiloGauss ($\omega \sim \frac{eH}{mc}\gamma $) is necessary.
Slow-wave devices (TWT, BWT, orotrons)in this range require
application of dense and thin (<0.1 mm) electron beams, because
only electrons passing near the slowing structure at the distance
$\leq \lambda \beta \gamma /(4\pi )$ can interact with
electromagnetic wave effectively.
It is difficult to guide thin beams near slowing structure with
desired accuracy. And electrical endurance of resonator limits
radiation power and density of acceptable electron beam.
Conventional waveguide systems are essentially restricted by the
requirement for transverse dimensions of resonator, which should
not significantly exceed radiation wavelength. Otherwise,
generation efficiency reduces abruptly due to excitation of plenty
of modes.

\qquad
The most of the above problems can be overpassed in VFEL.
In VFEL the greater part of electron beam interacts with electromagnetic wave
due to volume distributed interaction. Transverse dimensions of VFEL resonator
could significantly exceed radiation wavelength $D>>\lambda $. In addition,
electron beam and radiation power are distributed over the whole volume that is
beneficial for electrical endurance of the system.
Multi-wave Bragg dynamical diffraction provides mode discrimination in VFEL.

\qquad Dispersion equations, describing instability and threshold
conditions of generation for VFEL were investigated in details in
\cite{fer}-\cite{mm}. There were shown that threshold values of
current can be significantly reduced in conditions of multi-wave
Bragg diffraction. But there mainly relativistic and
ultrarelativistic electron beams were considered. Essential
distinctions appear at use of nonrelativistic and low-relativistic
electrons. For example, in slow-wave devices radiation conditions
can not be fulfilled simultaneously with conditions of Bragg
diffraction if refraction index $n<c/u$. Generation in
spatially-periodic structure composed from dielectric threads was
studied in \cite{exp} for low-relativistic (300 keV)electron beam.

\qquad Present paper investigates instability of electron beam in
volume diffraction structures composed from strained threads.
There are frequency ranges, in which such systems can generate,
and those in which they work as amplifiers. It should be mentioned
that generation can occur even in one-periodic diffraction
structure. In this case Bragg conditions are not fulfilled and
back-wave is used for generation. Tuning of radiation frequency in
such a system can be provided either by change of radiation angle
or by grating rotation or by change of direction and value of
electron beam velocity. Nonrelativistic and low-relativistic
electron beams passing through such structures can generate in
wide frequency range up to terahertz. \qquad

\section{Amplification and generation in diffraction structure}

\begin{figure}[tbp]
\epsfxsize =10cm \centerline{\epsfbox{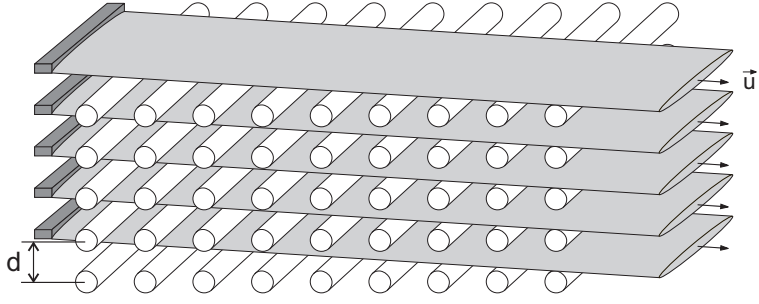}} 
\caption{ General view of Volume Free Electron Laser formed by
metal threads with several sheet electron beams.} \label{f1}
\end{figure}

\qquad Let us consider an electron beam with velocity $\vec{u}$
passing through a periodic structure composed from either
dielectric or metal threads (see Figure \ref{f1}.)
Fields, which appear at electron beam passing through a volume
spatially-periodic medium are described by the following set of
equations \cite{3}:

\begin{eqnarray}
DE-\omega ^{2}\chi _{1}E_{1}-\omega ^{2}\chi _{2}E_{2}-\omega ^{2}\chi
_{3}E_{3}-... &=&0  \notag \\
-\omega ^{2}\chi _{-1}E+D_{1}E_{1}-\omega ^{2}\chi _{2-1}E_{2}-\omega
^{2}\chi _{3-1}E_{3}-... &=&0  \label{system} \\
-\omega ^{2}\chi _{-2}E-\omega ^{2}\chi _{1-2}E_{1}+D_{2}E_{2}-\omega
^{2}\chi _{1-2}E_{1}-... &=&0,  \notag
\end{eqnarray}

Set (\ref{system})is obtained by the use of Bloch representation for field
\begin{equation*}
\vec{E}(\vec{r};\omega )=\sum \vec{E}_{i}\exp
\{i(\vec{k}+\vec{\tau}_{i})\vec{r}\} \end{equation*}
and permittivity in a spatially-periodic
medium
($\varepsilon
(\vec{r}+\vec{a}_{n},\omega )=\varepsilon (\vec{r},\omega )$)

\begin{equation*}
\varepsilon (\vec{r},\omega )=1+\sum_{\{\tau \}}\chi _{\tau }(\omega )\exp
(i\vec{\tau}\vec{r}),.
\end{equation*}
$\vec{\tau}_{i}=\{\frac{2\pi }{a}i_{1};\frac{2\pi
}{b}i_{2};\frac{2\pi }{c}i_{3}\}$ are the reciprocal lattice
vectors, $\vec{a}_{n}=n_{1}\vec{a}+n_{2}\vec{b}$ $+n_{3}\vec{c}$
are the translation vectors of periodic medium, $i_{1}\div i_{3}$,
$n_{1}\div n_{3}$ are the integer numbers. $D_{\alpha }=k_{\alpha
}^{2}c^{2}-\omega ^{2}\varepsilon +\chi _{\alpha }^{(b)}$ ,
$\vec{k}_{\alpha }=\vec{k}+\vec{\tau}_{\alpha }$ is the wave
vector of diffracted photon, $\chi _{\alpha }^{(b)}$ is the part
dielectric susceptibility, caused by the presence of electron
beam:

\begin{eqnarray}
\chi _{\alpha }^{(b)} &=&\frac{1}{\gamma }(\omega _{b}/\omega
)^{2}(\mathbf{ue}^{\alpha }/c)^{2}\times \frac{k_{\alpha }^{2}c^{2}-\omega ^{2}}{(\omega
-\vec{k}_{\alpha }\vec{u})^{2}}  \label{beam} \\
&&\text{ ''cold'' beam limit,}  \notag \\
\chi _{\alpha }^{(b)} &=&-\frac{i\sqrt{\pi }}{\gamma }(\omega _{b}/\omega
)^{2}(\mathbf{ue}^{\alpha }/c)^{2}\times \frac{k_{\alpha }^{2}c^{2}-\omega
^{2}}{\delta _{\alpha }^{2}}x_{\alpha }^{t}\exp [-(x_{\alpha }^{t})^{2}]
\notag \\
&&\text{''hot'' beam limit.} \notag
\end{eqnarray}%
$x_{\alpha }^{t}=(\omega -\vec{k}_{\alpha }\vec{u})/\sqrt{2}\delta _{\alpha
} $, $\delta _{\alpha }^{2}=(k_{\alpha 1}^{2}\Psi _{1}^{2}+k_{\alpha
2}^{2}\Psi _{2}^{2}+k_{\alpha 3}^{2}\Psi _{3}^{2})u^{2}$ and
$\vec{\Psi}=\Delta \vec{u}/u$ spread of electron's velocities in a beam.
Values of $\chi _{\alpha }^{(b)}$ in (\ref{beam}) are cited for two
opposite limits.
First case is described by the inequality
$|\omega -\vec{k}_{\alpha }\vec{u}|\gg \delta _{\alpha }$ and corresponds to the
so called hydrodynamic or ''cold'' beam limit.
In this case all electrons participate in interaction with electromagnetic wave.
Kinetic or ''hot'' beam limit $|\omega -\vec{k}_{\alpha
}\vec{u}|\precsim \delta _{\alpha }$ supposes that only part of electrons
participates in interaction process.

\qquad As it was mentioned above, synchronism conditions are
incompatible with conditions of dynamic diffraction for
nonrelativistic and low-relativistic electron beams. In this case
generation can appear due to transition radiation in periodic
medium  and electron beam interacts with slightly coupled
component corresponding to the wave vector $\vec{k}_{\tau
}=\vec{k}+\vec{\tau}$. According to (\ref{system}) the dispersion
equation, which describes instability of electron beam is as
follows:

\begin{equation}
(k^{2}c^{2}-\omega ^{2}\varepsilon )(k_{\tau }^{2}c^{2}-\omega
^{2}\varepsilon +\chi _{\tau }^{(b)})-\omega ^{2}\chi _{\tau }\chi _{-\tau
}=0.  \label{disp}
\end{equation}

As synchronism conditions are incompatible with Bragg those, then
$k^{2}\neq k_{\tau }^{2}$ in the instability range. At the same
time two different types of instability exist depending on
radiation frequency. Amplification takes place when electron beam
is in synchronism with electromagnetic component
$\vec{k}+\vec{\tau}$, which has positive projection $k_{z}$. If
projection $k_{z}$ is negative and generation threshold is
reached, then generation occurs. In the first case radiation
propagates along the transmitted wave, which has positive
projection of group velocity
$v_{z}=\frac{c^{2}k_{z}^{(0)}%
}{\omega }$ ($k_{z}^{(0)}=\sqrt{\omega ^{2}\varepsilon -k_{\perp
}^{2}}$), and beam disturbance moves along it. In the second case
the group velocity has negative projection
$v_{z}=-\frac{c^{2}k_{z}^{(0)}}{\omega }$ and radiation propagates
along back-wave and electromagnetic wave comes from the range of
the greatest beam disturbance to the place, where electrons come
into the interaction area. For one-dimensional structure such
mechanism is realized in backward-wave tube.

Dispersion equation, which describes the roots, corresponding interaction of
electron beam with electromagnetic wave at
$k^{2}\neq k_{\tau }^{2}$ can be rewritten as:

\begin{equation}
u_{z}^{2}\left( k_{z}-a_{1}\right)
^{2}(k_{z}-k_{z}^{(0)})(k_{z}+k_{z}^{(0)})=-\frac{a\omega
_{L}^{2}(\vec{u}\vec{e}^{\tau })^{2}\omega ^{4}r}{c^{4}\left( k_{\tau }^{2}c^{2}-\omega
^{2}\varepsilon _{0}\right) },  \label{slow}
\end{equation}
where $a_{1}=\frac{\omega -\vec{k}_{\perp }\vec{u}_{\perp
}}{u_{z}}-\tau _{z}$. For amplification case (\ref{slow}) gives for
increment of instability:

\begin{equation}
Im k_{z}^{\prime }=-\frac{\sqrt{3}}{2}f,  \label{increment}
\end{equation}
where $f=\sqrt[3]{\frac{a\omega _{L}^{2}(\vec{u}\vec{e}^{\tau
})^{2}\omega ^{4}r}{2k_{z}^{(0)}c^{4}u_{z}^{2}\left( k_{\tau
}^{2}c^{2}-\omega ^{2}\varepsilon _{0}\right) }}$, if condition
$2k_{z}^{\prime }f\gg \frac{\omega ^{2}\chi _{0}"}{c^{2}}$ is
fulfilled. In case $2k_{z}^{\prime }f\ll \frac{\omega ^{2}\chi
_{0}"}{c^{2}}$ dissipative instability evolves. Its increment is

\begin{equation}
Im
k_{z}=-\frac{c}{\omega}\sqrt{\frac{k_{z}^{(0)}f^{3}}{\chi_{0}}}.
\label{diss}
\end{equation}
If inequalities $k_{z}^{`2}\gg 2k_{z}k_{z}^{`}$ and $k_{z}^{`2}\gg
\frac{\omega ^{2}\chi _{0}"}{c^{2}}$ are fulfilled, spatial
increment of instability can be expressed as:

\begin{equation}
Im k_{z}^{`}=-\left( \frac{a\omega _{L}^{2}(\vec{u}\vec{e}^{\tau
})^{2}\omega ^{4}r}{c^{4}\left( k_{\tau }^{2}c^{2}-\omega
^{2}\varepsilon _{0}\right) u_{z}^{2}}\right) ^{1/4}, \label{four}
\end{equation}
but parameters providing dependence
(\ref{four})
correspond to the conversion from amplification to generation regime.
For Compton instability (at proper electron beam density) this situation
takes place at $k_{z}^{(0)}\approx 0$

\qquad Frequency of amplified radiation is defined as:

\begin{equation}
\omega =\frac{\vec{\tau}\vec{u}}{1-\beta _{x}\eta _{x}-\beta _{y}\eta
_{y}-\beta _{z}\sqrt{\varepsilon -\eta _{x}^{2}-\eta _{y}^{2}}}.
\label{freqa}
\end{equation}

\begin{figure}
\epsfxsize =10cm \centerline{\epsfbox{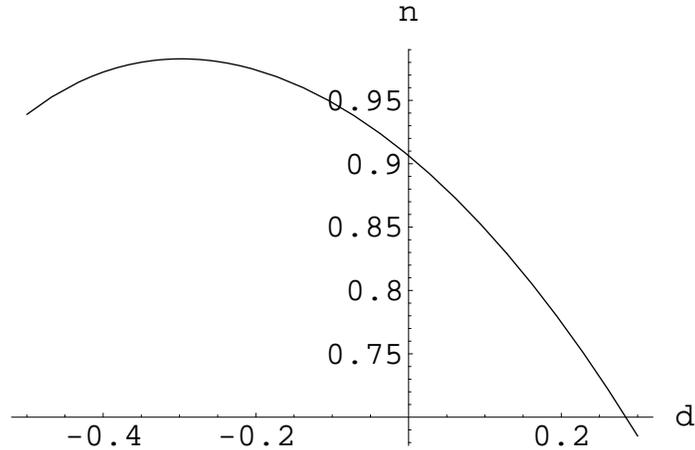}}
\caption{ Calculated dependence of temporal increment on detuning
parameter.} \label{Fig.1}
\end{figure}

\qquad Instability in generation regime
is described by the temporal increment and
can not be described by the spatial
that. Increment of absolute instability can be found solving the equation

\begin{equation}
Im k_{z}^{(+)}(\omega )=Im k_{z}^{(-)}(\omega ) \label{condition}
\end{equation}

with respect to imaginary part of $\omega $. Roots of (\ref{slow})
are expressed as:

\begin{gather}
k_{z1}^{(-)}=a_{1}-\frac{a_{1}+k_{z}^{(0)}}{3}+A+B  \label{cardano} \\
k_{z2,3}^{(+)}=a_{1}-\frac{a_{1}+k_{z}^{(0)}}{3}-\frac{A+B}{2}\pm
i\frac{A-B}{2}\sqrt{3},  \notag
\end{gather}
where

\begin{gather}
A=\sqrt[3]{\frac{f^{3}}{2}-\left( \frac{a_{1}+k_{z}^{(0)}}{3}\right)
^{3}+\sqrt{\frac{f^{6}}{4}-f^{3}(a_{1}+k_{z}^{(0)})^{3}}}  \notag \\
B=\sqrt[3]{\frac{f^{3}}{2}-\left( \frac{a_{1}+k_{z}^{(0)}}{3}\right)
^{3}-\sqrt{\frac{f^{6}}{4}-f^{3}(a_{1}+k_{z}^{(0)})^{3}}}.  \notag
\end{gather}
Calculated dependence of temporal increment on parameter of
detuning is presented in Fig.\ref{Fig.1}.

Axes in Fig.\ref{Fig.1} are denoted as:
$d=\frac{a_{1}+k_{z}^{(0)}}{f}$, $n=\frac{\frac{\omega
"}{u_{z}}+\frac{2\omega \omega "\varepsilon _{0}+\omega ^{2}\chi
_{0}"}{c^{2}k_{z}^{(0)}}}{f}$.
It follows from Fig.\ref{Fig.1} that
at certain
value of parameter of detuning:
\begin{equation}
\frac{a_{1}+k_{z}^{(0)}}{f}\approx -0.3
\label{detuning}
\end{equation}
increment of instability has maximum peak
\begin{equation}
\frac{\frac{\omega "}{u_{z}}+\frac{2\omega \omega "\varepsilon _{0}+\omega
^{2}\chi _{0}"}{c^{2}k_{z}^{(0)}}}{f}\approx 0.98.  \label{solution}
\end{equation}
Increment of absolute instability can be found from (\ref{solution}).
It is easy to see that absolute instability can evolve if current exceed
start value, which is determined by dissipation.
Amplification regime has no threshold and deciding influence of dissipation causes
dissipative instability (\ref{diss}).

\qquad Frequencies, corresponding to generation regime, are
defined by the expression different from (\ref{freqa}):

\begin{equation}
\omega =\frac{\vec{\tau}\vec{u}}{1-\beta _{x}\eta _{x}-\beta _{y}\eta
_{y}+\beta _{z}\sqrt{\varepsilon -\eta _{x}^{2}-\eta _{y}^{2}}}.
\label{freqr}
\end{equation}

\qquad Thus, it follows from (\ref{freqa}, \ref{freqr}) that
change of radiation angle causes smooth frequency tuning. At that,
generation frequencies are less than those corresponding to
amplification regime. Hence, using system as amplifier one should
add dispersion elements in it to raise dissipation in frequency
range, in which generation occurs.

 Use of Bragg multiwave distributed feedback increases generation
 efficiency and provides discrimination of generated modes.
 If conditions of synchronism and Bragg conditions are not fulfilled
 simultaneously, diffraction structures with two different periods can be applied
\cite{nonrel}. The first of them provides synchronism of
electromagnetic wave with electron beam $\omega
-\vec{k}\vec{u}=\vec{\tau}_{1}\vec{u}$, where $\vec{\tau}_{1}$ is
the reciprocal lattice vector of this structure. The second
diffraction structure evolves distributed Bragg coupling
$|\vec{k}|\approx |\vec{k}+\vec{\tau}_{j}|$, $\vec{\tau}_{j}$
($j=2\div n$)are the reciprocal lattice vectors of the second
structure. Conditions of synchronism and Bragg conditions can be
fulfilled even for one diffraction structure, because diffraction
structure with one period has unlimited set of reciprocal lattice
vectors. Then reciprocal lattice vectors providing synchronism and
Bragg conditions are significantly different in value.

\qquad It follows from (\ref{system}) that dispersion equation for
distributed feedback supplied by two-wave diffraction with one low
coupled wave is expressed as:

\begin{equation}
F_{3}(\vec{k};\vec{k}_{1};\vec{k}_{2})=-\chi
_{1}^{(b)}F_{2}(\vec{k};\vec{k}_{2})  \label{three}
\end{equation}

where:

\begin{eqnarray*}
F_{3}(\vec{k};\vec{k}_{1};\vec{k}_{2}) &=&(k^{2}c^{2}-\omega ^{2}\varepsilon
)(k_{1}^{2}c^{2}-\omega ^{2}\varepsilon )(k_{2}^{2}c^{2}-\omega
^{2}\varepsilon )- \\
&&\omega ^{4}(k^{2}c^{2}-\omega ^{2}\varepsilon )\chi _{1-2}\chi
_{2-1}-\omega ^{4}(k_{1}^{2}c^{2}-\omega ^{2}\varepsilon )\chi _{2}\chi _{-2}
\\
&&-\omega ^{4}(k_{2}^{2}c^{2}-\omega ^{2}\varepsilon )\chi _{1}\chi
_{-1}-\omega ^{6}(\chi _{1}\chi _{-2}\chi _{2-1}+\chi _{2}\chi _{-1}\chi
_{1-2}) \\
F_{2}(\vec{k};\vec{k}_{2}) &=&(k^{2}c^{2}-\omega ^{2}\varepsilon
)(k_{2}^{2}c^{2}-\omega ^{2}\varepsilon )-\omega ^{4}\chi _{2}\chi _{-2}.
\end{eqnarray*}
(\ref{three}) is derived with the assumption that synchronism
conditions are fulfilled for reciprocal lattice vector
$\vec{\tau}_{1}$ ($\omega -(\vec{k}+\vec{\tau}_{1})\vec{u}\approx
0$), while two-wave diffraction evolves at planes with reciprocal
lattice vector $\vec{\tau}_{2}$ ($|\vec{k}|\approx
|\vec{k}+\vec{\tau}_{2}|$). Threshold conditions of generation and
increment of temporal instability for the latter geometry were
obtained in \cite{nonrel}:

\begin{equation}
\omega ^{\prime \prime }=\frac{\omega }{2(1-\beta )}\{G-(\frac{\gamma
_{0}u}{\vec{n}\vec{u}})^{3}\frac{16\pi ^{3}n^{2}}{-\beta (k\chi _{2}L_{\ast
})^{2}kL_{\ast }}-\chi _{0}^{\prime \prime }(1-\beta \pm \sqrt{-\beta
}\frac{r^{\prime \prime }}{|\chi _{2}|\chi _{0}^{\prime \prime }})\}
\label{threshold}
\end{equation}
where:

\begin{eqnarray*}
G &=&(kL_{\ast })^{2}\frac{\pi ^{2}n^{2}}{4\gamma }(\frac{\omega
_{b}}{\omega })^{2}(\frac{\vec{u}\vec{e}_{1}}{\vec{n}\vec{u}})^{2}(l_{1}+\chi _{0})
\\
&&\times \frac{l\chi _{1-2}\chi _{2-1}+l_{2}\chi _{1}\chi _{-1}+\chi
_{1}\chi _{-2}\chi _{2-1}+\chi _{2}\chi _{-1}\chi _{1-2}}{l_{1}^{2}}f(y),
\end{eqnarray*}%
$l=\frac{k_{0}^{2}c^{2}-\omega ^{2}\varepsilon }{\omega ^{2}}$,
$l_{1}=\frac{(\vec{k}_{0}+\vec{\tau}_{1})^{2}c^{2}-\omega ^{2}\varepsilon }{\omega
^{2}}$, $l_{2}=\frac{(\vec{k}_{0}+\vec{\tau}_{2})^{2}c^{2}-\omega ^{2}\varepsilon
}{\omega ^{2}}.$
Condition $\omega
^{\prime \prime }=0$ in (\ref{threshold}) defines the start current of generation.
Threshold conditions for s-wave diffraction is converted to:

\begin{equation}
G^{(s)}=\frac{a_{s}^{3}}{(k\chi L_{\ast })^{2s}kL_{\ast }}+\chi _{0}^{\prime
\prime }b_{s}.  \label{s_wave}
\end{equation}
For developed dynamic diffraction,
when $k|\chi |L_{\ast }\gg 1$,
either generation start current
or
length of generation zone at certain current value
can be reduced.

\qquad
Each Bragg condition holds one of free parameters.
For example,
for certain geometry and electron beam velocity
two conditions for 3-wave diffraction entirely determine
transverse components of wave vector
$k_{x}$ and $k_{y}$, and therefore generation frequency
(see (\ref{freqa}, \ref{freqr})).
Hence, volume diffraction system provides
mode discrimination due to multiwave diffraction.

\section{\qquad\  Discussion.}

\qquad
The above results affirm that volume diffraction structure provides both
amplification and generation regimes even in the absence of dynamic diffraction.
In latter case generation evolves with backward-wave similarly
backward-wave tube.
Frequency in such structure is changed smoothly either at smooth
variation of radiation angle (variation of $k_x$ and $k_y$) or at
rotation of diffraction grating or electron beam (change of
$\vec{\tau}\vec{u}$)(see (\ref{freqa}), (\ref{freqr})). For
certain geometry and reciprocal lattice vector amplification
corresponds higher frequencies then generation. Rotation of either
diffraction grating or electron beam also changes value of
boundary frequency, which separates generation and amplification
ranges. Use of multiwave distributed feedback owing to Bragg
diffraction, let either to increase generation efficiency or to
reduce length of interaction area ((\ref{threshold}),
(\ref{three})). In this case generation is available with both
backward and following waves.

\qquad
In particular, the proposed volume structure can be used for generation of
sub-millimeter radiation by accelerator
 LIU-3000.
Parameters of this setup:
electron beam energy
 $E=800$ keV, beam current $I=100\div 200$ A.
To generate radiation with wavelength 0.3 mm in such a system
volume structure composed from strained threads should has
period $\sim $ 2 mm, and period of diffraction grating providing Bragg
coupling is $\sim $ 0.16 mm.


\bigskip

\bigskip


\begin{references}
\bibitem{fer}  V.G.Baryshevsky, I.D.Feranchuk Phys.Let {\bf 102A},141,(1984).

\bibitem{2}  V.G.Baryshevsky, K.G.Batrakov, I.Ya.Dubovskaya Journ.Phys.D
{\bf 24},1250,(1991).


\bibitem{3}  V.G.Baryshevsky, K.G.Batrakov, I.Ya.Dubovskaya NIM {\bf A 358},
 493,(1995).

\bibitem{mm}  V.G.Baryshevsky, K.G.Batrakov, I.Ya.Dubovskaya, S.Sytova NIM A
{\bf 358},508,(1995).

\bibitem{exp} V.G.Baryshevsky, K.G.Batrakov, I.Ya.Dubovskaya, V.A.Karpovich,
V.N.Rodionova, NIM A {\bf 393}, p. 71-75, {1997}.

\bibitem{nonrel} V.G.Baryshevsky, K.G.Batrakov, V.I.Stolyarsky
Proceedings of 21 FEL Conference, p. 37-38, {1999}.


\end{references}
\end{document}